\documentclass{article}
\usepackage{graphicx}
\title{\textbf{Canonical wave packets in quantum cosmology}}
\author{S. S. Gousheh, H. R. Sepangi, P. Pedram\thanks{Email: pedram@sbu.ac.ir}, M. Mirzaei, \\ {\small Department of Physics,
Shahid Beheshti University, Evin, Tehran 19839, Iran}}
\begin{document}
\maketitle \baselineskip 24pt

\begin{abstract}
We discuss the construction of wave packets resulting from the
solutions of a class of  Wheeler-DeWitt equations in
Robertson-Walker type cosmologies, for arbitrary curvature. We
show that there always exists a  ``canonical initial slope" for a
given initial wave function, which optimizes some desirable
properties of the resulting wave packet, most importantly good
classical-quantum correspondence. This can be properly denoted as
a canonical wave packet. We introduce a general method for finding
these canonical initial slopes which is generalization of our
earlier work.
\vspace{5mm}
\end{abstract}

\section{Introduction}\label{I}
The question of the construction and interpretation of  wave
packets in quantum cosmology and its connection with classical
cosmology has been attracting much attention in recent years.
Moreover, numerous studies have been done to obtain a quantum
theory for gravity and to understand its connection with classical
physics. The problem of the relation between quantum cosmology and
classical physics is an important one that exists even in simple
models \cite{pad}. Most authors consider semi-classical
approximations to the Wheeler-DeWitt (WD) equation and refer to
regions in configuration space where these solutions are
oscillatory or exponentially decaying as representing classically
allowed or forbidden regions, respectively. These regions are
mainly determined by the initial conditions for the wave function.
Two popular proposals for the initial conditions are  the {\it no
boundary proposal} \cite{hawking} and the {\it spontaneous
nucleation from nothing} \cite{vilenkin}. These proposals have
been  attractive to many authors because they lead to some classes
of classical solutions represented by certain trajectories which
posses important features such as predicting an inflationary
phase.

In quantum cosmology, in analogy with ordinary quantum mechanics,
one is generally concerned with the construction of wave functions
by the superposition of the `energy eigenstates' which would peak
around the classical trajectories in configuration space, whenever
such classical-quantum correspondence is mandated by the nature of
the problem. However, contrary to ordinary quantum mechanics, a
parameter describing time is absent in quantum cosmology. Therefore,
the initial conditions would have to be expressed in terms of an
{\it intrinsic} time parameter, which in the case of the WD equation
could be taken as the local scale factor for the three geometry
\cite{dewitt}. Also, since the sign of the kinetic term for the
scale factor is negative, a formulation of the Cauchy problem for
the WD equation is possible. The existence of such a sign is one of
the exclusive features of gravity with many other interesting
implications.

The construction of wave packets resulting from the solutions of
the WD equation has been a feature common to most of the recent
research work in quantum cosmology \cite{kiefer,tucker,sepangi}.
In particular, in reference \cite{kiefer} the construction of wave
packets in a Friedmann universe is presented in detail and
appropriate boundary conditions are motivated. Generally speaking
one of the aims of these investigations has been to find wave
packets whose probability distributions coincide with the
classical paths obtained in classical cosmology. In these works,
the authors usually consider model theories in which a self
interacting scalar field is coupled to gravity in a
Robertson-Walker type universe with zero $k$. The resulting WD
equation is often in the form of an isotropic
oscillator-ghost-oscillator and is separable in the configuration
space variables. The general solution can thus be written as a sum
over the product of simple harmonic oscillator wave functions with
different frequencies. As usual, the coefficients in the sum and
hence the exact form of the wave packets are determined by the
initial conditions, which are usually specified by giving the wave
function and its derivative at $R=0$, where $R$ is the scale
factor. The coefficient are chosen such that the following
desirable properties are obtained. Firstly, there should be a good
classical-quantum correspondence, which means the wave packet
should have compact support centered around the classical path,
the crest of the wave packet should follow as closely as possible
the classical path, and to each distinct classical path there
should correspond a wave packet with the above properties. The
first part of this condition implies that the initial wave
function should consist of a few localized pieces. Secondly, one
expects the square of the wave packet describing a physical system
to posses a certain degree of smoothness.

At the classical level, if the initial position is known, the
corresponding momentum can be calculated from the constraint
$H=0$. However, in quantum cosmology the WD equation can be in
general put in the form of a second order hyperbolic PDE.
Therefore, complete description of the initial condition would in
general entail the specification of both the initial wave function
and its first derivative. However, it is in general an unresolved
problem as to how, by knowing the initial classical values of the
positions and momenta, one may embark on obtaining the
corresponding quantum distributions. Generally speaking, there is
no generic way of doing this. What is sometimes done is to
consider a distribution which peaks around the classical values of
the variables of interest, mostly and for simplicity taken as a
Gaussian distribution, and simply let the resulting wave packet
evolve and hope for a good classical quantum correspondence
\cite{dabrowski,ashtekar}.  Indeed, even if the wave packet does
follow the classical path it might not be smooth. Smooth wave
packets have been constructed in the past by adapting coherent
states for the initial wave function and the undetermined
coefficient related to the initial first derivative distribution
are chosen to have the same functional form as those of the
initial wave function \cite{wavepacket}. Here, we shall generalize
the aforementioned prescription to be applicable to any reasonable
potential in the same class of problems. Moreover, the
prescription is general enough to be applicable to problems which
are not exactly solvable. The situation is therefore becomes
analogous to the classical case: Given an initial position
distribution, there is a canonical initial momentum distribution
which ensures good classical-quantum correspondence. This is in
contrast to the usual practice where there is no uniform
prescription for choosing initial momentum distribution in order
to obtain a good classical quantum correspondence as possible is
obtained. Our solution here is based on \cite{wavepacket} where a
certain prescription was offered for obtaining initial canonical
slope.

The paper is organized as follows: In Sec.~\ref{II} we outline the
main problem which is a case of Robertson-Walker type cosmology
where the matter is taken to be a particular type of
self-interacting scalar field. We derive the main equations both
for the classical cosmology and the quantum one. In
Sec.~\ref{III}, we present a general prescription for finding a
Canonical Initial Slope (CIS) for a given choice of suitable
initial condition on the wave function, which produces the above
desired properties of the wave packet. In this derivation we will
come up with a decisive criteria for what initial wave functions
are suitable for classical-quantum correspondence. This
prescription is general enough to be applicable to both exactly
solvable cases and the ones which are not. Therefore, we begin
Sec.~\ref{III} with a description the Spectral Method (SM)
\cite{pedram} which is a robust numerical method.  We then present
the $k=0$ case which is exactly solvable and display the resulting
wave packet obtained by CIS. Then we consider the case $k\ne 0$,
for which neither the classical nor the quantum equations of
motion are exactly solvable. In particular, in our model the
classical equations of motion turn out to be a set of non-linear
coupled ODEs with moving singularities. For the classical cases,
we solve them by a numerical method that we introduced in an
earlier work \cite{siamak2}. In the quantum cosmology case we
solve the problem using SM. For this case we display the resulting
wave packets obtained by the CIS prescription. In Sec.~\ref{IV} we
draw some final conclusions.

\section{Robertson-Walker cosmology with a coupled scalar field}\label{II}
In Robertson-Walker cosmology, one often considers
the coupling of a scalar field  to gravity. The resulting field
equations include a `zero energy constraint'. The WD equation in
quantum cosmology is the result of the quantization consistent with
this constraint. Now we discuss our model and derive the field
equations in the classical and quantum domains.

Consider the Einstein's field equation coupled to a scalar field
\begin{eqnarray}
R_{\mu\nu}-\frac{1}{2}g_{\mu\nu}R&=&\kappa T_{\mu\nu}(\phi), \label{eq1} \\
\Delta^2\phi-\frac{\partial U}{\partial \phi}&=&0, \label{eq2}
\end{eqnarray}
where $T_{\mu\nu}$ is the energy-momentum tensor coupling the
scalar field $\phi$ to gravity and $U$ is a scalar potential for
the scalar field $\phi$. Hereafter, we shall work in a system of
units in which $\kappa=1$. We parameterize the metric as
\begin{eqnarray}
g=-dt^2+R^2(t)\frac{\sum dr^i dr^i}{(1+kr^2/4)^2},  \label{eq3}
\end{eqnarray}
where $R(t)$ is the usual scale factor and $k=+1,0,-1$ corresponds
to a closed, flat or open universe, respectively. The field
equations resulting from (\ref{eq1}) and (\ref{eq2}) with the
metric given by (\ref{eq3}) can be written as
\begin{eqnarray}
3\left[\left(\frac{\dot{R}}{R}\right)^2+\frac{k}{R^2}\right]&=&
\frac{\dot{\phi}^2}{2}+U(\phi), \label{eq4}\\
2\left(\frac{\ddot{R}}{R}\right)+\left(\frac{\dot{R}}{R}\right)^2+
\frac{k}{R^2}&=&-\frac{\dot{\phi}^2}{2}+U(\phi), \label{eq5}\\
\ddot{\phi}+3\frac{\dot{R}}{R}\dot{\phi}+\frac{\partial U}{\partial\phi}
&=&0.  \label{eq6}
\end{eqnarray}
Here a dot represents differentiation with respect to time. We
require the potential $U(\phi )$ to have natural characteristics for
small $\phi$, so that we may identify the coefficient of
$\frac{1}{2}\phi^2$ in its Taylor expansion as a positive mass
squared $m^2$, and $U(0)$ as a cosmological constant $\Lambda$. An
interesting choice of $U(\phi)$ with three free parameters is
\cite{tucker,darabi1,wavepacket,darabi2,ahmadi},
\begin{eqnarray}
U(\phi)=\Lambda+\frac{m^2}{2\alpha^2}\sinh^2(\alpha\phi)+
\frac{b}{2\alpha^2}\sinh(2\alpha\phi).  \label{eq7}
\end{eqnarray}
In the above expression $\Lambda$ may be identified with a
cosmological constant, $m^2=\partial^2 U/\partial\phi^2|_{\phi=0}$
is a mass squared parameter, $b$ is a coupling constant, and
$\alpha^2=\frac{3}{8}$. However, one of our main reasons for
choosing this potential is that its exact solutions exist for both
the classical and quantum cosmology for $k=0$ \cite{wavepacket} and
we want to solve the problem for $k\ne0$ and compare the solutions
to the corresponding ones for $k=0$. To do this, we have general
numerical methods that can solve both the classical \cite{siamak2}
and quantum cosmological \cite{pedram} cases. More importantly,
having the exact solution for $k=0$ and approximate solution for
$k\ne 0$ for the same potential simultaneously, we can formulate a
general procedure for finding the CIS and test it. Our claim would
then be the resulting general formulation works for any potential in
the same category. This gives us motivation to solve this problem
for $k\ne0$ for the case of quantum cosmology. The Lagrangian giving
the above equations of motion can be written as
\begin{eqnarray}
L=-3R\dot{R}^2+3kR+R^3 [\dot{\phi}^2/2-U(\phi)].
\label{eq8}
\end{eqnarray}
This Lagrangian can be cast into a simple form by using the
transformations $X=R^{3/2} \cosh(\alpha\phi)$ and $Y=R^{3/2}
\sinh(\alpha\phi)$, which transform the term $R^3 U(\phi )$ into
a quadratic form. Upon using a second transformation to eliminate
the coupling term in the quadratic form, we arrive at new
variables $u$ and $v$, which are linear combinations of $X$ and
$Y$,
\begin{eqnarray}
\left(
\begin{array}{l}
 u \\
 v
\end{array}
\right)=\left(
\begin{array}{rr}
 \cosh(\theta) & -\sinh(\theta) \\
 -\sinh(\theta) & \cosh(\theta)
\end{array}
\right)\left(
\begin{array}{l}
 X \\
 Y
\end{array}
\right)
\end{eqnarray}
where,
\begin{eqnarray}
\theta=\frac{1}{2} \tanh ^{-1}\left(\frac{-2 b}{m^2}\right).
\end{eqnarray}
In the terms of the new variables, the Lagrangian takes on the
following simple form,
\begin{eqnarray}
L(u,v)=\frac{4}{3}\left[ (\dot{u}^2-\omega_1^2
u^2)-(\dot{v}^2-\omega_2^2 v^2)-\frac{9}{4}k (u^2-v^2)^{1/3}
\right],\label{eq9}
\end{eqnarray}
where $\omega_{1,2}^2 = -3\lambda /4 +m^2/2 \mp\sqrt{m^4-4b^2}/2$.
The resulting classical field equations are,
\begin{eqnarray}
\ddot{u}+\omega_1^2u+\frac{3k}{2}\frac{u}{(u^2-v^2)^{2/3}}=0,\label{u''}
\end{eqnarray}
\begin{eqnarray}
\ddot{v}+\omega_2^2v+\frac{3k}{2}\frac{v}{(u^2-v^2)^{2/3}}=0,\label{v''}
\end{eqnarray}
\begin{eqnarray}
\dot{u}^2+\omega_1^2u^2-\dot{v}^2-\omega_2^2v^2+\frac{9}{4}k(u^2-v^2)^{1/3}=0.\label{u2}
\end{eqnarray}
Equations (\ref{u''}) and (\ref{v''}) are the dynamical equations
and (\ref{u2}) is the zero energy constraint. The non-linearity and
the moving singular behavior of these equations are now apparent.

The corresponding quantum cosmology is described by the
Wheeler-DeWitt (WD) equation written as,
\begin{eqnarray}
H\psi(u,v)=\left\{-\frac{\partial^2}{\partial
u^2}+\frac{\partial^2} {\partial v^2}+ \omega_1^2 u^2-\omega_2^2
v^2+\frac{9}{4}k (u^2-v^2)^{1/3}\right\}\psi(u,v)=0, \label{eq10}
\end{eqnarray}
which arises from the zero energy condition Eq.\ (\ref{u2}).

\section{Solutions for the classical and quantum cosmology cases}\label{III}
In this section we shall present a general prescription for
finding CIS. The cases that we shall study will include some cases
which are exactly solvable ($k=0$) and some which are not
($k\ne0$). For the latter we have to resort in part to some
numerical methods. Therefore, we start this section by a
discussion of the numerical method that we shall use, so that
later on we can concentrate on the physics of the problem with no
interruption. The general PDE that we want to solve is
$$\left\{-\frac{\partial^2}{\partial u^2}+\frac{\partial^2}
{\partial v^2}+ \hat{f}'(u,v)\right\}\psi(u,v)=0,$$ where
$\hat{f}'(u,v)$ is an arbitrary function. This equation has the
general form of a hyperbolic PDE equation and is encountered in
many branches of science and in particular physics. It is notable
that such equations may represent a wave-like equation whose
solution may rapidly oscillate. In such cases, the usual spatial
integration routines such as Finite Difference Methods fail to
produce a reasonable solution. It is therefore of prime importance
to introduce a method of solution which is efficient, accurate,
predictable and easily implemented, given any reasonable boundary
conditions \cite{pedram}.

Equation (\ref{eq10}) in general is not separable except when
$k=0$ \cite{wavepacket}, and we have to  resort to a numerical
method to solve this hyperbolic PDE. Among the various numerical
techniques, we choose SM which has the following advantages: It is
very simple, fast, accurate, robust and stable. Spectral Method
(SM) \cite{SP1,SP2,SP3}, consists of first choosing a complete
orthonormal set of eigenstates of a, preferably relevant,
hermitian operator to be used as a suitable basis for our
solution. For this numerical method we obviously cannot choose the
whole set of the complete basis, as these are usually infinite.
Therefore we make the approximation of representing the solution
by only a finite superposition of the basis functions. By
substituting this approximate  solution into the differential
equation, a matrix equation is obtained. The expansion
coefficients of these approximate solutions could be determined by
eigenfunctions of this matrix. In this method, the accuracy of the
solution is increased by choosing a larger set of basis functions.
Having resorted to a numerical method, it is worth setting up a
more general problem defined by the following WD equation,
\begin{eqnarray}
H\psi(u,v)=\left\{-\frac{\partial^2}{\partial
u^2}+\frac{\partial^2} {\partial v^2}+ \omega_1^2 u^2-\omega_2^2
v^2+\hat{f}(u,v)\right\}\psi(u,v)=0. \label{eq10new}
\end{eqnarray}
As mentioned before, any complete orthonormal set can be used. In
this section we use the Fourier series basis by restricting the
configuration space to a finite square region of sides $2L$. This
means that we can expand the solution as,
\begin{eqnarray}
\psi(u,v)=\sum_{i,j=1}^2 \sum_{m,n} A_{m,n,i,j} \,\,\,
g_i\left(\frac{m \pi u}{L}\right)\,\,\, g_j\left(\frac{n \pi
v}{L}\right), \label{eqpsitrigonometric}
\end{eqnarray}
where,
\begin{equation}\label{eqsincos}
\left\{
  \begin{array}{ll}
    g_1\left(\frac{m \pi
u}{L}\right)=\sqrt{\frac{2}{R_{m} L}}\sin\left(\frac{m \pi
u}{L}\right), &  \\
   g_2\left(\frac{m \pi
u}{L}\right)=\sqrt{\frac{2}{R_{m} L}}\cos\left(\frac{m \pi
u}{L}\right). & \\
  \end{array}
\right. \mbox{and}\,\,\, R_{m} =\left\{ \begin{array}{ll}
    1, & m\ne0 \\
    2, & m=0 \\
\end{array}
\right.
\end{equation}
By referring to the WD equation (\ref{eq10new}), we realize that in
the Fourier basis it is appropriate to introduce $\hat f'$ as,
\begin{eqnarray}
\hat f'(u,v)=\hat{f}(u,v)+\omega_1^2 u^2-\omega_2^2
v^2.\label{eqf'}
\end{eqnarray}
We can make the following expansion,
\begin{eqnarray}\label{eqf'2}
\hat f'(u,v) \psi(u,v)=\sum_{i,j} \sum_{m,n} B'_{m,n,i,j} \,\,\,
g_i\left(\frac{m \pi u}{L}\right)\,\,\, g_j\left(\frac{n \pi
v}{L}\right).
\end{eqnarray}
where $B'_{m,n,i,j} $ are coefficients that can be determined once
$\hat f'(u,v)$ is specified. By substituting
(\ref{eqpsitrigonometric},\ref{eqf'2}) in (\ref{eq10new}), and using
independence of $g_i\left(\frac{m \pi u}{L}\right)$s and
$g_j\left(\frac{n \pi v}{L}\right)$s we obtain,
\begin{eqnarray}
\left[\left(\frac{m \pi }{L}\right)^2 -\left(\frac{n \pi
}{L}\right)^2\right] A_{m,n,i,j}+B'_{m,n,i,j}=0,\label{eqAB'}
\end{eqnarray}
where,
\begin{eqnarray}
B'_{m,n,i,j}\hspace{-3mm}&=& \sum_{m',n',i',j'}
\left[\int\hspace{-2mm}\int_{-L}^{L} g_{i}\left(\frac{m \pi
u}{L}\right) g_{j}\left(\frac{n \pi v}{L}\right) \hat f'(u,v)
g_{i'}\left(\frac{m' \pi u}{L}\right) g_{j'}\left(\frac{n' \pi
v}{L}\right)du dv\right]A_{m',n',i',j'}\nonumber\\ &=&
\sum_{m',n',i',j'} C'_{m,n,i,j,m',n',i',j'}\,\,
A_{m',n',i',j'}.\label{B'}
\end{eqnarray}
Therefore we can rewrite (\ref{eqAB'}) as
\begin{eqnarray}
\left[\left(\frac{m \pi}{L} \right)^2 -\left(\frac{n \pi}{L}\right
)^2\right] A_{m,n,i,j}+ \sum_{m',n',i',j'}
C'_{m,n,i,j,m',n',i',j'}\,\, A_{m',n',i',j'}=0.\label{eqAC'}
\end{eqnarray}
In this case, we select $4N^2$ basis functions, that is $m$ and $n$
run from $1$ to $N$. It is obvious that the presence of the operator
$\hat f'(u,v)$ leads to nonzero coefficients
$C'_{m,n,i,j,m',n',i',j'}$ in (\ref{eqAC'}), which in principle
could couple all of the matrix elements of $A$. Then we replace the
square matrix $A$ with a column vector $A'$ with $(2N)^2$ elements,
so that any element of $A$ corresponds to one element of $A'$. This
transforms (\ref{eqAC'}) to
\begin{eqnarray}
D\, A'=0. \label{eqmatrix2}
\end{eqnarray}
Matrix $D$ is a square matrix now with $(2N)^2 \times (2N)^2$
elements which can be easily obtained from (\ref{eqAC'}). Equation
(\ref{eqmatrix2}) can be looked upon as an eigenvalue equation, {\it
i.e.} $DA'_a=a A'_a$ and the matrix $D$ has $(2N)^2$ eigenvectors.
However, for constructing the acceptable wavefunctions, {\em i.e.
}the ones satisfying the WD equation (\ref{eq10new}), we only
require eigenvectors which span the null space of the matrix $D$.
That is, due to (\ref{eqAC'}) we will have exactly $2N$ null
eigenvectors which will be linear combination of our original
eigenfunctions introduced in (\ref{eqpsitrigonometric}). After
finding the $2N$ eigenvectors of $D$ with zero eigenvalue, {\em
i.e.} $A'^k$ ($ k=1,2,3,...,2N$), we can find the corresponding
elements of matrix $A$, $A^k_{m,n,i,j}$. Therefore, the wavefunction
can be expanded as
\begin{eqnarray}
\psi(u,v)= \sum_{k} \lambda^k \psi^k(u,v)=\sum_{k} \lambda^k
\sum_{m,n,i,j} A^k_{m,n,i,j}\,\,\, g_i\left(\frac{m \pi
u}{L}\right)\,\,\, g_j\left(\frac{n \pi v}{L}\right).
\label{eqpsifinal2}
\end{eqnarray}
Here $\lambda^k\,\,$s in (\ref{eqpsifinal2}) are arbitrary complex
constants to be determined by the initial conditions.

We are free to adjust two parameters: $2N$, the number of basis
elements used and the length of the spatial region, $2L$. This
length should be preferably larger than spatial spreading of all the
sought after wave functions. However, if $2L$ is chosen to be too
large we loose overall accuracy. Therefore it is important to note
that for each $N$, $L$ can be properly adjusted \cite{SP1}. 

The most important physical aspects of this problem that we want
to address is on the question of the existence of a CIS. In our
previous investigation of the same problem for $k=0$
\cite{wavepacket} we discovered that a ``canonical" choice for the
slope exists whose use produces wave packets with all the desired
properties stated in the introduction. Here we have come up with a
more general and systematic method for obtaining this slope. The
results are based on the same general prescription as before and
obviously give identical results for comparable cases. This slope
turns out to depend on the nature of the hamiltonian and the
initial wave function. Since the WD equation is a hyperbolic PDE,
for any reasonable choice of initial wave packet, there will
remain certain undetermined coefficients which will be pending the
specification of the initial first derivative. Our prescription is
simply to choose the undetermined coefficient to have the same
functional form as the determined ones. We shall clarify this
prescription further in some concrete examples. If the problem is
exactly solvable this slope can be easily obtained by the method
described in Ref. \cite{wavepacket}. If it is not, we can use the
following prescription:  We use the minimum approximation
necessary to make the general PDE (\textit{i.e.} (\ref{eq10new}))
separable near $v=0$ and solve the resulting equations. It is
obvious that the presence of the odd terms of $v$ dose not have
any effect on the form of the initial wave function but they are
responsible for the slope of the wave function at $v=0$, and vice
versa for the even terms. We can approximate (\ref{eq10}) near the
$v=0$, so up to the first order in $v$ we have,
\begin{eqnarray}
\left\{-\frac{\partial^2}{\partial u^2}+\frac{\partial^2}
{\partial v^2}+ \omega_1^2 u^2+\frac{9}{4}k
(u^2)^{1/3}\right\}\psi(u,v)=0. \label{eq10nearv0}
\end{eqnarray}
This PDE is separable in $u$ and $v$ variables, so we can write,
\begin{equation}\label{psi-separated}
\psi(u,v)=\varphi(u)\chi(v).
\end{equation}
By using this definition in (\ref{eq10nearv0}), two ODEs can be
derived,
\begin{eqnarray}
\frac{d^2\chi_n(v)}{dv^2}+E_n\chi_n(v)&=&0,
\label{eqseparated1}\\
\hspace{-0.6cm}-\frac{d^2\varphi_n(u)}{d u^2}+\left(\omega_1^2
u^2+\frac{9}{4}ku^{2/3}\right)\varphi_n(u)&=&E_n\varphi_n(u),\label{eqseparated2}
\end{eqnarray}
where $E_n$s are separation constants. These equations are
Schr\"{o}dinger-like equations with $E_n$s as their `energy'
levels. Equation (\ref{eqseparated1}) is exactly solvable with
plane wave solutions or,
\begin{equation}\label{eqplanewave}
\chi_n(v)=\alpha_n\cos\left(\sqrt{E_n}\,\,v\right)+i\beta_n\sin\left(\sqrt{E_n}\,\,v\right),
\end{equation}
where $\alpha_n$ and $\beta_n$ are arbitrary complex numbers.
Equation (\ref{eqseparated2}) does not seem to be exactly solvable
and we resort to a numerical technique. As mentioned before, SM can
be used to find the bound state energy levels ($E_n$) and the
corresponding wave functions ($\varphi_n(u)$) with high accuracy.
The general solution to the (\ref{eq10nearv0}) can be written as,
\begin{equation}\label{psi-separated2}
\psi(u,v)=\sum_{n=\mbox{\footnotesize{even}}} (A_n
\cos(\sqrt{E_n}v)+iB_n\sin(\sqrt{E_n}v))
\varphi_n(u)+\sum_{n=\mbox{\footnotesize{odd}}}( C_n
\cos(\sqrt{E_n}v)+iD_n\sin(\sqrt{E_n}v)) \varphi_n(u).
\end{equation}
The separation of this solution to even and odd terms, though in
principle unnecessary, is crucial for our prescription for the CIS.
As stated before, this solution is valid only for small $v$. The
general initial conditions can now be written as,
\begin{eqnarray}
\psi(u,0)=\sum_{even}A_n\varphi_n(u)+\sum_{odd}C_n\varphi_n(u)\label{eqinitial1}\\
\hspace{-0.8cm}\psi'(u,0)=i\sum_{even}B_n\sqrt{E_n}\,\,\varphi_n(u)+i\sum_{odd}D_n\sqrt{E_n}\varphi_n(u),\label{eqinitial2}
\end{eqnarray}
 where prime denotes the derivative with respect to $v$.
Obviously a complete description of the problem would include the
specification of both these quantities. However, given only the
initial condition on the wave function, we claim there is only a
CIS which produces a canonical wave packet with all the
aforementioned desired properties. We can qualitatively describe
our prescription for this case as setting the functional form of
the odd undetermined coefficients to be the same as the even
determined ones and vice versa. This means that the coefficients
that determine CIS i.e. $B_n$ for $n$ even and $D_n$ for $n$ odd,
are chosen by the following,
\begin{equation}\label{eqcanonicalslope}
  B_n=C_n\,\,\, \,\,\,\,\mbox{for $n$ even}\hspace{1cm}D_n=A_n\,\,\,\,\,\, \mbox{for $n$
  odd}
\end{equation}
To be more specific, although $C_n$ ($A_n$) is defined only for
$n$ odd (even), we can extend its definition to $n$ even (odd) by
choosing the same functional form.

Now we want to discuss the general settings for quantum-classical
correspondence. As mentioned before in the classical cosmology
case, the initial condition is given by specifying a few isolated
initial points in the configuration space and uniquely  determined
velocities associated with each one through the constraint $H=0$.
In order to established a good quantum-classical correspondence,
the first thing we should do is to specify an initial wave
function with isolated peaks centered around the classical ones.
Then we can use the CIS prescription to obtain the initial first
derivative distribution. One of the main outcomes of our
prescription is that the expectation value of initial momentum
computed using our prescription in the quantum cosmology case,
denoted by $\left[\langle p_v \rangle \right]_{v=0}$, is
approximately the same as the corresponding classical quantity
$\left[\dot{v}(t)\right]_{t=0}$ (in our notation) for each
isolated classical point. That is $\left[\langle p_v \rangle
\right]_{v=0}\rightarrow \left[\dot{v}(t)\right]_{t=0}$ where,
\begin{equation}\label{eqavemomentum}
\hspace{-0.1cm} \left[\langle p_v\rangle
\right]_{v=0}=\int\left[\psi(u,v)^*
\left(-i\frac{\partial}{\partial
    v}\right)\psi(u,v)\right]_{v=0}\hspace{-0.5cm}du.
\end{equation}
Obviously the domain of the above integral should extend only over
the appropriate part of the initial wave function, which
corresponds to the classical point in question. This relationship
approaches an equality for initial states for which classical
description is appropriate. We shall shortly make this statement
more explicit for the particular problem under investigation in
this paper. Our prescription for the CIS is general enough to be
applicable to any initial wave function.

\begin{figure*}[b]
\centerline{\begin{tabular}{ccc}
\includegraphics[width=8cm]{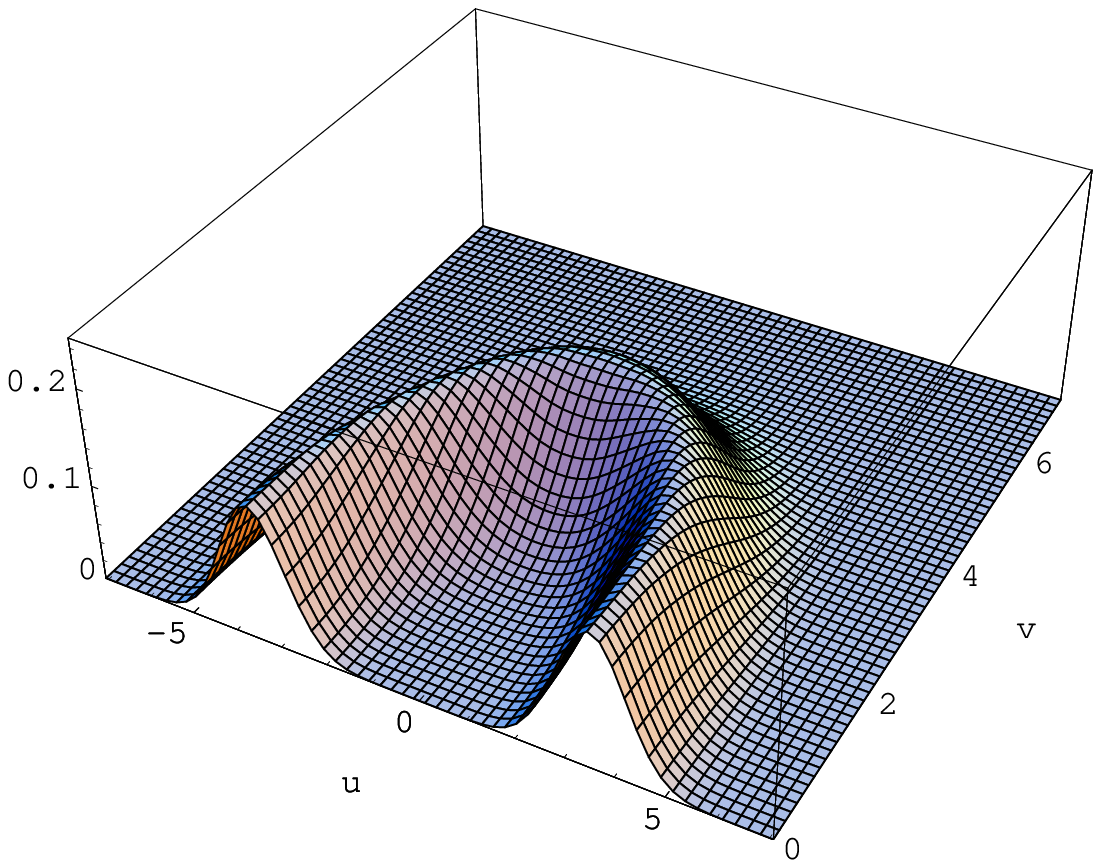}
 &\hspace{2.cm}&
\includegraphics[width=6.5cm]{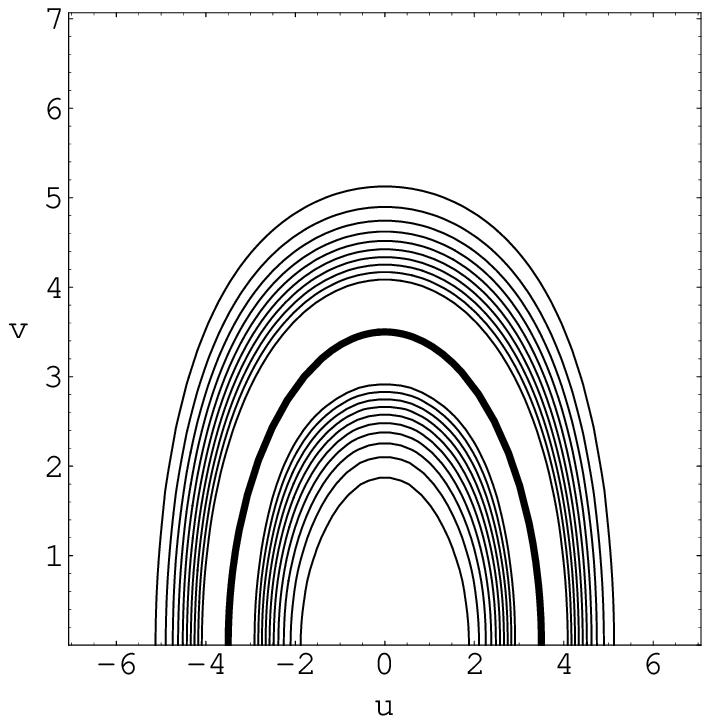}
\end{tabular}}
\caption{$k=0$ case: Left, the square of the wave packet $|
\psi(u,v)|^2$ for $\chi=3.5$ and $N=15$. Right, the contour plot
of the same figure with the classical path superimposed as the
thick solid line.} \label{fig1}
\end{figure*}

In the case $k=0$ the problem is exactly solvable \cite{wavepacket}
and has a closed form solution in the Simple Harmonic Oscillator
(SHO) basis. In references \cite{kiefer,wavepacket} the authors
considered the following initial condition for the wave function,
\begin{equation}
\psi(u,0)=\frac{1}{2
\pi^{1/4}}\left(e^{-(u-\chi)^2/2}+e^{-(u+\chi)^2/2}\right).\label{eqpsi0}
\end{equation}
which consists of two symmetric coherent states of a one dimensional
SHO. Note that, for the given choice of the initial wave function we
can not expect a good classical-quantum correspondence for $\chi<3$,
since there would be a significant overlap between the two pieces,
which could cause quantum interference. With this choice for the
initial wave function one might expect the classical-quantum
correspondence to be manifest for $\chi>3$. However, as we shall
see, this crucially depends on the choice of the initial slope.
Obviously, this choice for $\psi(u,0)$ is not unique and is chosen
so that the resulting wave packet would have compact support. In the
case $k=0$, (\ref{eq10}) is exactly separable without any
approximation and yields exactly the same Schr\"{o}dinger-like
equation, i.e. SHO, as the one obtained from (\ref{eqseparated2})
for both $u$ and $v$ variables with frequencies $\omega_1$ and
$\omega_2$ respectively. The solutions to these equation are well
known, i.e. for $u$ is
\begin{eqnarray}\label{eqk0}
\varphi_n^0(u)=\left(\frac{\omega_1}{\pi}\right)^{1/4}\frac{H_n(
\sqrt{\omega_1}u)} {\sqrt{2^n n!}}e^{-\omega_1 u^2/2},\hspace{2cm}
E_n^0=(2n+1)\omega_1,
\end{eqnarray}
where $H_n\,$s are the Hermite polynomials and superscript zero
indicates $k=0$. Then the complete eigenstates will be products of
eigenstates for $u$ and $v$, with constraint that they should have
the same energy. In this case the initial wave function is chosen
to be even and taking into account our prescription for the CIS
(\ref{eqcanonicalslope}), (\ref{eqinitial1},\ref{eqinitial2}) can
be rewritten as,
\begin{eqnarray}
\psi(u,0)&=&\sum_{even}c_n\varphi_n^0(u)\label{eqinitial1k0}\\
\psi'(u,0)&=&i\sum_{odd}c_n\sqrt{(2n+1)\omega_1}\,\,\varphi_n^0(u),\label{eqinitial2k0}
\end{eqnarray}
where the choice,
\begin{eqnarray}
c_n=e^{-\frac{1}{4} |\chi|^2}\frac{\chi^n}{\sqrt{2^n n!}}\, ,
\label{eqcn}
\end{eqnarray}
produces our aforementioned wave function (\ref{eqpsi0}) and $\chi$
is an arbitrary complex number. The CIS introduced in Ref.
\cite{wavepacket}, based on the same prescription as introduced here
but for the exactly solvable case $k=0$, is,
\begin{eqnarray}
\psi'(u,0)=\sum_{n_{{\mbox{\tiny odd}}}} \frac{c_{n}
\varphi_{n}^0(u)H'_n(0)}{\frac{(-1)^{(n/2)} n!}{(n/2)!}},
\label{eqinitial3k0}
\end{eqnarray}
where prime denotes differentiation with respect to $v$. Numerical
comparison between (\ref{eqinitial2k0}) and (\ref{eqinitial3k0})
shows that their relative difference is ${\mathcal O}\,(10^{-4})$.
Their difference is due to the fact that (\ref{eqinitial3k0}) is
exact and (\ref{eqinitial2k0}) is approximate. We can now easily
construct the coherent wave packets using (\ref{eqinitial1k0}) and
(\ref{eqinitial2k0}) or (\ref{eqinitial3k0}). For illustrative
purposes we choose the simplest case possible which is when
$\omega_1=\omega_2$. The result is shown in the left part of
Fig.~\ref{fig1}. As can be seen from this figure the parameters of
the problem are chosen such that the initial state consists of two
well separated peaks and this class of problems are the ones which
are also amenable to a classical description. We will make this
statement more quantitative at the end of this section. We should
mention that there are a variety of different cases illustrated in
Ref.~\cite{wavepacket} for the same exactly solvable case of
$k=0$, including $\omega_1\ne \omega_2$ obviously using
(\ref{eqinitial3k0}).

\begin{figure*}[b]
\centerline{
\includegraphics[width=8cm]{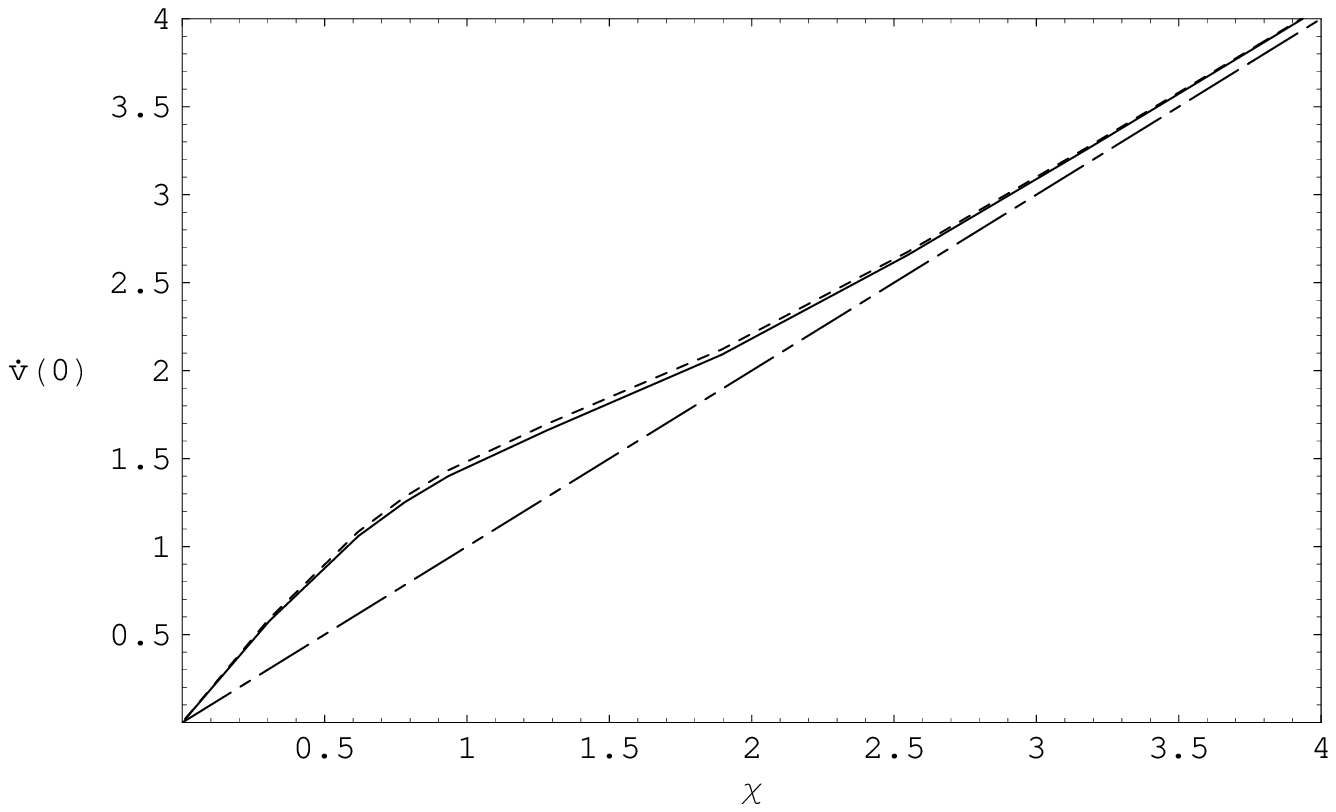}}
\caption{$\dot{v}(0)$ versus $\chi$ for the three cases: classical
(dashed-dot), exact CIS (dashed), and approximate CIS (solid). On
this scale the latter two approximately coincide for $\chi > 4$.
Their maximum difference is of order $10^{-2}$. These two CISs
asymptotically approach the classical value for $\chi > 8$.)}
\label{fig10}
\end{figure*}

The classical paths corresponding to these solutions can be easily
obtained from (\ref{u''},\ref{v''}). The corresponding initial
conditions for the classical case is the following:
\begin{eqnarray}
 u(0)=-\chi,\hspace{0.5cm} v(0)=0,\hspace{0.5cm}
\dot{u}(0)=0,\hspace{0.5cm}
\dot{v}(0)=\dot{v}_0,\label{eqclassicalic}
\end{eqnarray}
and the parameters $\chi$ and $\dot{v}_0$ are adjusted so that the
zero energy condition (\ref{u2}) is satisfied. With this choice of
initial conditions the exact classical paths would be, in the case
$k=0$, Lissajous figures in general, which are circles with radii
$\chi$ when $\omega_1=\omega_2\equiv \omega$. Using
(\ref{eqavemomentum}) we can compute the average initial momentum
in the quantum case for the left peak, where the corresponding
classical values are $u(0)=-\chi$, $v(0)=0$, $\dot{u}(0)=0$ and
$\dot{v}=+\omega\chi$. It is important to note that the domain of
integration in equation (\ref{eqavemomentum}) in this case should
be from $-\infty$ to zero, as explained before. We obtain the
following two expressions first using the exact value for CIS
(\ref{eqinitial3k0}), and second the approximate CIS
(\ref{eqinitial2k0}),
\begin{eqnarray}\label{eqavemomentum2}
    \left[\langle p_v \rangle \right]^{\mbox{\footnotesize{exact}}}_{v=0}&=&4 \sum_{m_{{\mbox{\tiny odd}}}}\sum_{n_{{\mbox{\tiny even}}}}\frac{c_n c_m }{\sqrt{2^n n! \sqrt{\pi}}\sqrt{2^m m! \sqrt{\pi}}}
    \,\,\left[\frac{(m/2)!\sqrt{\omega}H'_m(0)}{(-1)^{((m+1)/2)} m!}\right]\,\,n!m!\sum_{k=0}^{\tiny{\mbox{Min}\{m,n\}}}\frac{2^k
    H(0)_{m+n-2k-1}}{k!(n-k)!(m-k)!}\\
    \left[\langle p_v \rangle\right]^{\mbox{\footnotesize{approx.}}}_{v=0}&=&4 \sum_{m_{{\mbox{\tiny odd}}}}\sum_{n_{{\mbox{\tiny even}}}}\frac{c_n c_m }{\sqrt{2^n n! \sqrt{\pi}}\sqrt{2^m m! \sqrt{\pi}}}
    \,\,\left[i\sqrt{(2m+1)\omega}\right]\,\,n!m!\sum_{k=0}^{{\mbox{\tiny Min}\{m,n\}}}\frac{2^k H(0)_{m+n-2k-1}}{k!(n-k)!(m-k)!}
\end{eqnarray}
Having precisely set the initial conditions for both the classical
and quantum cosmology cases, we can now superimpose the results
for the case $k=0$, as illustrated in the right part of the Fig.\
\ref{fig1}. As can be seen in the figure, and also for all the
cases presented in \cite{wavepacket}, the classical-quantum
correspondence is manifest.
In Fig. \ref{fig10} we show $\dot{v}(0)$ versus $\chi$ for the
three cases: classical, exact CIS, and approximate CIS. As can be
seen from this figure the classical-quantum correspondence is
meaningful only for $\chi > 3$, that is where there is no
significant overlap between the two pieces of $\psi(u,0)$. This
correspondence becomes better as $\chi$ increases. That is
$\left[\langle
p_v\rangle\right]_{v=0}=\left[\dot{v}(t)\right]_{t=0}=\dot{v}_0$
for large $\chi$. We can generically expect loss of
classical-quantum correspondence when there is significant quantum
interference between different segments of the wave packet in the
configuration space, when each segment is supposed to correspond
to a distinct classical configuration. We would like to emphasize
that the choice of CIS is crucial for establishing a good
correlation. For example in Ref. \cite{kiefer} the initial slope
in the quantum case is chosen to be zero, implying zero initial
average momentum (see (\ref{eqavemomentum})), however the
corresponding classical quantity $\dot{v}(0)=\pm\omega\chi\neq 0$
and the classical path would again be a circle. To be more
precise, the WD equation emanates from the zero energy condition
and their choice of initial conditions violates this constraint.
Therefore it would be impossible to establish a classical-quantum
correspondence in this case. It is interesting to note that their
resulting wave packet ($|\psi(u,v)|^2$) is wildly oscillating.

\begin{figure*}[b]
\centerline{\begin{tabular}{ccc}
\includegraphics[width=8cm]{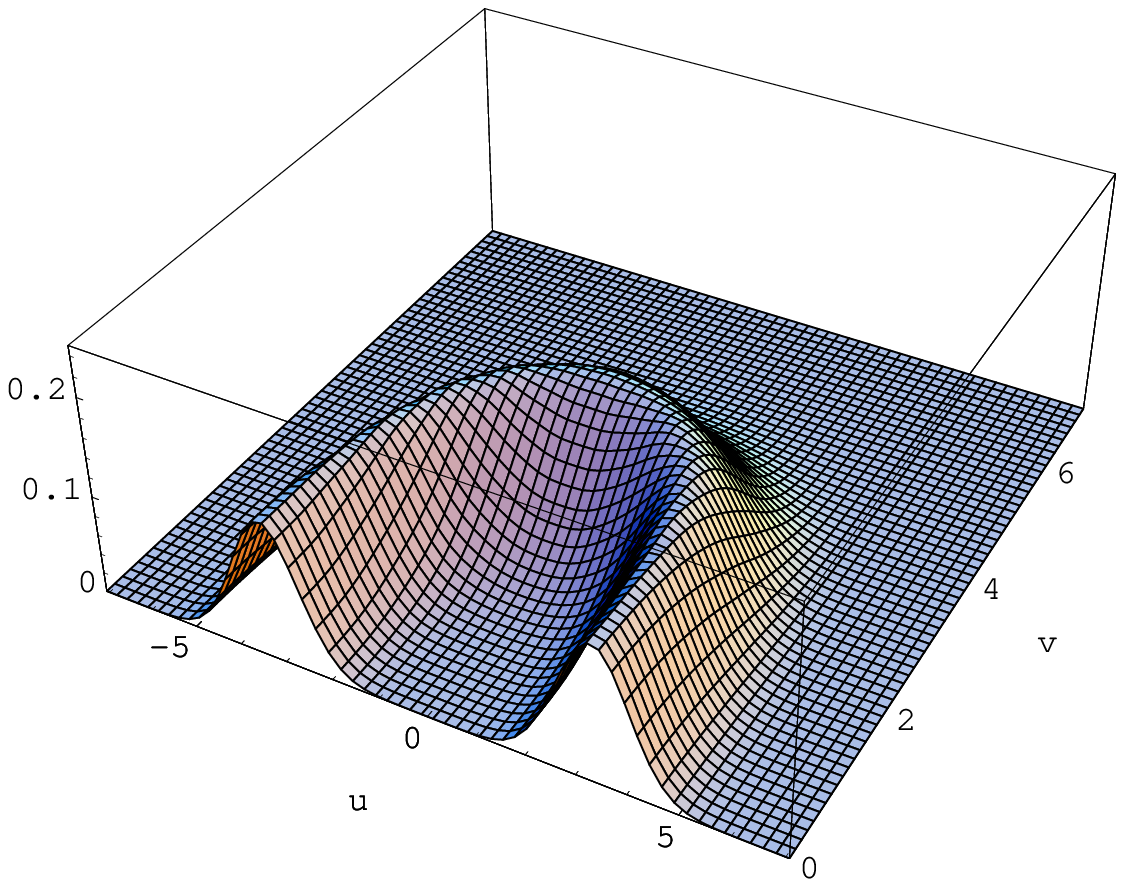}
 &\hspace{2.cm}&
\includegraphics[width=6.5cm]{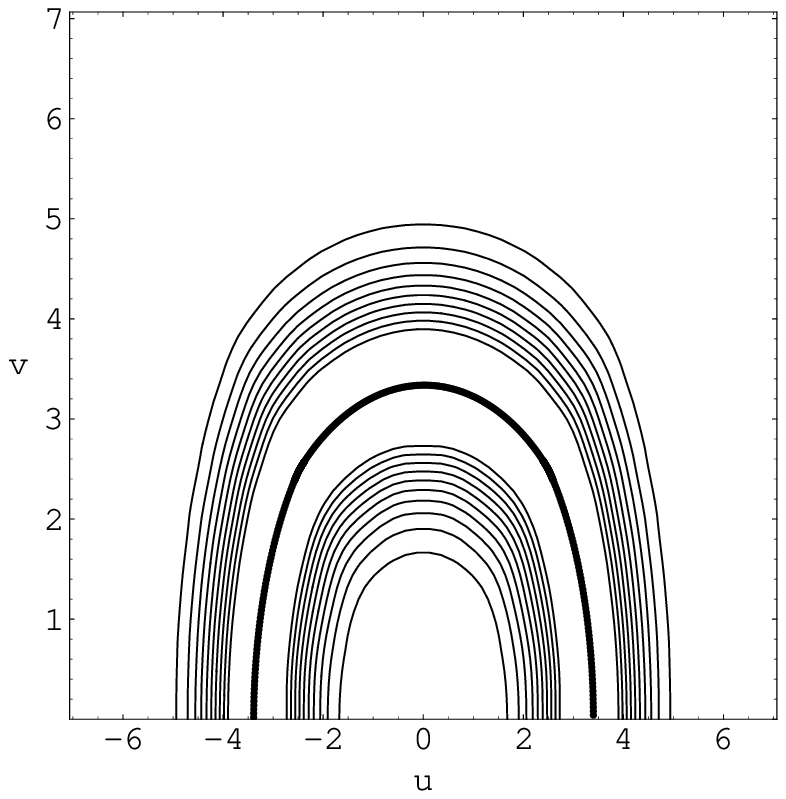}
\end{tabular}}
\caption{$k=1$ case: Left, the square of the wave packet $|
\psi(u,v)|^2$ for $\chi=3.5$ and $N=15$. Right, the contour plot
of the same figure with the classical path superimposed as the
thick solid line.} \label{fig2}
\end{figure*}

\begin{figure*}[b]
\centerline{\begin{tabular}{ccc}
\includegraphics[width=8cm]{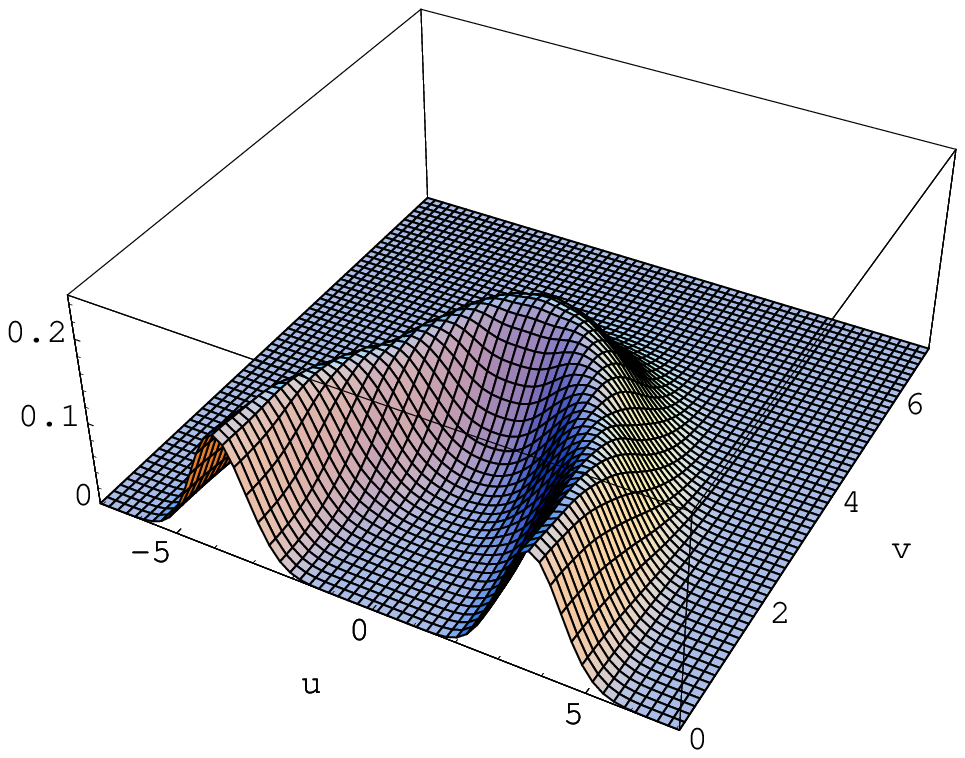}
 &\hspace{2.cm}&
\includegraphics[width=6.5cm]{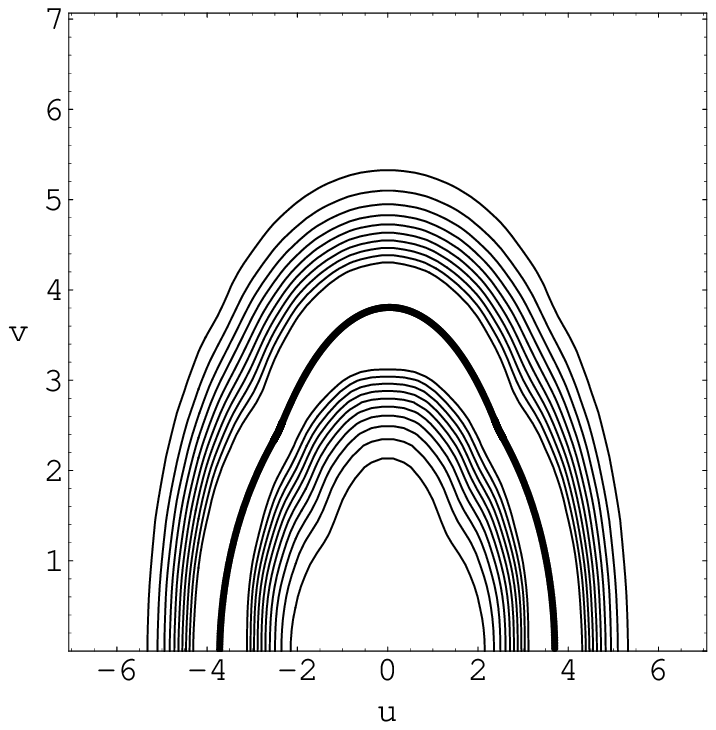}
\end{tabular}}
\caption{$k=-1$ case: Left, the square of the wave packet $|
\psi(u,v)|^2$ for $\chi=3.5$ and $N=15$. Right, the contour plot
of the same figure with the classical path superimposed as the
thick solid line.} \label{fig3}
\end{figure*}

\begin{figure*}[b]
\centering
\includegraphics[width=10cm]{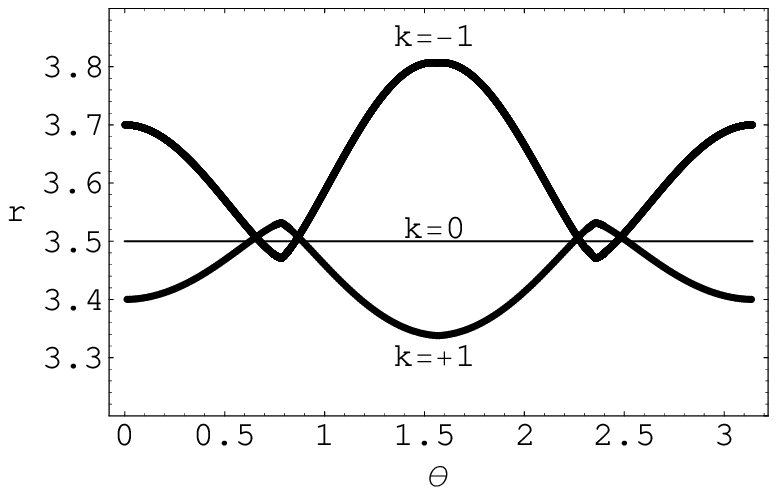}\\
\caption{Classical paths for $k=\{1,0,-1\}$, where
$r=\sqrt{u^2+v^2}$ and $\theta=\arctan(v/u)$.}\label{fig4}
\end{figure*}

For the case $k\ne 0$, the problem is not exactly solvable in
quantum cosmology and we will use the SM to get an approximate
solution. We have to mention that the corresponding equations for
classical cosmology, (\ref{u''})-(\ref{u2}), are a set of
nonlinear, coupled ODEs with moving singularities which are not
exactly solvable either. However, a general method for solving
them has been presented in \cite{siamak2}, and this is the method
we shall use. Also a detailed explanation of the physical setting
of the problem in the classical domain has been presented in
\cite{siamak}. For ease of comparison, we choose the same
illustrative problem with $\omega_1=\omega_2=\omega$ and the same
coefficients for the initial wave functions as the case $k=0$
(i.e. (\ref{eqinitial1k0},\ref{eqcn})). However, note that since
the eigenstates $\phi_n^{\pm 1}(u)$ and $\phi_n^{0}(u)$ (both
obtainable from (\ref{eqseparated2})) are slightly different, the
resulting initial wave functions for $k=\pm 1$ would be slightly
different from (\ref{eqpsi0}). From
(\ref{eqseparated2},\ref{eqinitial2}) it is apparent that the CISs
of the wave functions are different in all cases ($k=0,1,-1$), due
to the differences both in $\phi_n(u)\,$s and $E_n\,$s. We have
computed both the classical and quantum solutions for the
$k=+1,-1$ cases and the results are shown in Figs.~\ref{fig2} and
\ref{fig3}, respectively. Note that the wave packets are very
smooth and we have good classical-quantum correspondence for both
cases. Moreover, we have computed the graphs of $\dot{v}(0)$
versus $\chi$ for $k=\pm 1$ and have found the classical-quantum
correspondence to be as good as that exhibited in Fig. \ref{fig10}
for $k=0$.
In order to highlight the differences between the classical paths
(and obviously the wave packets)  we exhibit $r=\sqrt{u^2+v^2}$
versus $\theta=\arctan(v/u)$ in Fig. \ref{fig4} for the three cases.
\section{Conclusions}\label{IV}
We have described a Robertson-Walker type cosmology leading to
classical dynamical equations given by (\ref{u''}-\ref{u2}), and
the corresponding WD equation represented by (\ref{eq10}). All
these equations are exactly solvable for the particular potential
chosen when $k=0$ \cite{wavepacket}, and do not seem to be so when
$k \ne 0$. In the latter cases we have solved these equations
numerically: the classical cosmology equations by the numerical
method introduced in \cite{siamak2}, and the quantum cosmology
case by an implementation of the SM \cite{pedram}. Most
importantly, we have introduced a general method for finding the
canonical initial slopes for any initial wave function and any
value of $k$, whose use produces wave packets which can be named
canonical wave packets. The latter have the following desirable
properties: The correlation part of classical-quantum
correspondence is acceptable, as outlined in detail in  the
Introduction. Moreover, the resulting wave packets are very smooth
and this is usually considered to be a desirable property. Our
investigation has also revealed that the classical-quantum
correspondence is possible only when there is no overlap between
the lumps of a suitably chosen initial wave function. Using the
CIS, the average initial momenta become very close to their
corresponding classical values in the limit where the overlap
between the separate parts of the initial wave function goes to
zero. It is important to note that our generalized method is also
applicable to cases which are not exactly solvable.

\section*{Acknowledgement}
This research has been supported by the office of
research of Shahid Beheshti University under Grant No. 500/3787.

\end{document}